\documentstyle[preprint,aps,eqsecnum]{revtex}

\begin{document}

\title{Specific Heat of a Three Dimensional Metal Near a $T=0$ Magnetic
       Transition with Dynamic Exponent $z=2,3$, or $4$}
\author{U.\ Z\"ulicke and A.\ J.\ Millis}
\address{AT\&T Bell Laboratories, 600 Mountain Ave, Murray Hill, NJ 07974, USA}

\maketitle

\begin{abstract}
We derive expressions for the universal contribution to the
specific heat of a three-dimensional metal near a zero-temperature phase
transition with dynamic exponent $z=2,3$, or 4. The results allow a
quantitative comparison of theory to data. We illustrate the application of our
results by analyzing data for Ce$_{1-x}$Lu$_x$Cu$_2$Si$_2$, which has been
claimed to be near a quantum critical point.
\end{abstract}

\newpage

\narrowtext

\section{Introduction}
Many strongly-correlated metallic compounds
\cite{review} exhibit "non-fermi-liquid" behavior in the sense that the low
temperature ($T$) specific heat ($C$) exhibits more rapid temperature
dependence
than expected from the standard expression for a fermi liquid with fermi
temperature $T_F$, which is
\begin{equation}\label{norfersh}
\frac{C}{T} = \frac{A}{T_F} \left( 1 + B \frac{T^2}{{T_F}^2} \log T + {\cal O}
(\frac{T^2}{T_{F}^2}) + \dots \right)
\end{equation}
with $A=\frac{k_B\pi^2}{2}\, n$ ($n$ denotes the electronic density) and $B$
given in terms of Landau parameters \cite{normferm}. Many authors have
suggested
that the non-fermi-liquid behavior is due to the proximity of the
non-fermi-liquid materials to a zero temperature critical point
\cite{andy,continent,tsvelik,exotic}. One class of proposed critical points
would separate a $T=0$ magnetically ordered phase from a phase with no long
range order \cite{andy,continent,tsvelik}; another class involves critical
points occuring in models of isolated impurities \cite{exotic}. To
experimentally distinguish these intriguing suggestions it is necessary to
compare the data to theory. In this communication we provide quantitative
expressions for the universal contributions to the specific heat of a three
dimensional metal near a zero temperature magnetic---non-magnetic phase
transition with a dynamical exponent $z=2,3$, or 4. The case $z=2$ describes
antiferromagnetic transitions not driven by $2k_F$ instabilities or nesting;
the
case $z=3$ describes ferromagnetic transitions in a clean metal and also
describes the recently discussed "gauge theories" \cite{gauge} of strongly
correlated systems; the case $z=4$ would describe a dirty ferromagnet in the
temperature regime in which the randomness leads to spin diffusion but does not
otherwise change the critical properties.

Our results are obtained by solving scaling equations
derived previously \cite{andy}. The scaling equations imply that the critical
behavior is in the Gaussian universality class, with a dangerously irrelevant
operator. The results depend on three dimensionless parameters: reduced
temperature $t=\frac{T}{T^*}$ (where $T^*$ is a microscopic temperature
presumably of the order of the fermi temperature $T_F$), a control parameter
$r$
(which depends on temperature and on a Hamiltonian parameter such as pressure
or
doping whose variation tunes the material through the $T=0$ transition, which
is
defined to occur at $r(T=0)=0$) and $u$, the coefficient of the "dangerously
irrelevant operator". The results of course also depend on the spatial
dimensionality $d$ and the dynamical exponent $z$, and the results we write are
only valid for $d+z>4$. A qualitative phase diagram is shown in Fig.1. The axes
are reduced temperature , $t$, and the zero-temperature control parameter
$\delta=r(T=0)$. A phase transition line $t_c(\delta)$ separates a magnetically
ordered phase (shaded region) from a phase with no long range order. We give
results for the specific heat coefficient $\gamma=C/T$ in the disordered phase.
The behavior in the ordered phase depends on details of the order. Too many
special cases arise to discuss here. In the disordered phase we find that the
singular part, $\gamma_{sing}$, of $\gamma=\frac{C}{T}$ may be written
\begin{equation}\label{singpart}
\gamma_{sing} = \frac{k_B}{T^*} t^{\frac{d-z}{z}}\;
g_{d,z}\left(\frac{r}{t^{2/z
}},\frac{u t^{\frac{d+z-2}{z}}}{r}\right)
\end{equation}
where $g_{d,z}$ is a universal scaling function which we calculate. The leading
behavior of $\gamma_{sing}$ is given by $\frac{k_B}{T^*} t^{\frac{d-z}{z}}\,
g_{d,z}\left(\frac{r}{t^{2/z} },0\right)$ and may be computed directly from the
Gaussian model. The function $g_{d,z}(\frac{r}{t^{2/z}},0)$ describes the
crossover between a low-temperature (so-called quantum) regime with $t\ll r^{
\frac{z}{2}}$ and a high-temperature (so-called classical) regime defined by
$t\gg r^{\frac{z}{2}}$. A generally valid expression for $g_{3,z}(x,0)$ is
given
in Eq.\ (\ref{gaussheat2}). Expansions for the quantum and classical limits are
given in Eqs.\ (\ref{start})--(\ref{finish}). We also investigate the
dependence
of $g_{d,z}(x,y)$ on its second argument and show that it leads to "corrections
to scaling", which are then given as a power series in $u$ (times appropriate
combinations of $t$ and $r$). The only important correction to scaling arises
from the nonanalyticity in $g_{d,z}(x,y)$ occurring at $y=\infty$, i.e.\ $r=0$.
The equation $r(t,\delta,u)=0$ defines the true transition curve $t_c(\delta)$.
A precise expression is given in Eq.\ (\ref{3Dtransi}). To an extremely good
approximation, this transition is a second order mean field transition and is
characterized by a specific heat jump which we calculate. The result is Eq.\
(\ref{3Djump}). We also calculate the Gaussian fluctuation correction to this
specific heat jump. This is given in Eq.\ (\ref{3Ddiverge}).

The rest of this paper is organized as follows. In Section \ref{formal} we
write
down the model to be solved and indicate the method of solution. In Section
\ref{results} we give the results for $d=3$ and $z=2,3,4$. In Section
\ref{experi} we illustrate the comparison of the results to data \cite{csjee}
by comparing $g_{3,2}$ to data on Ce$_{1-x}$Lu$_x$Cu$_2$Si$_2$, a
three-dimensional material which is argued to be near a $T=0$ antiferromagnetic
transition. Section \ref{conclusion} is a brief conclusion. Details of various
calculations are given in Appendices.

\section{Formalism}\label{formal}
We follow the previous approach \cite{andy,hertz} of integrating out the
fermionic degrees of freedom and writing the partition function as a
functional integral over an N-component bosonic order parameter $\vec\phi$
which
represents the ordering field. This procedure is not useful for superconducting
transitions, transitions at nesting wavevectors of nested fermi surfaces, or
indeed transitions at wavevectors $Q$ connecting two points on the fermi
surface
with parallel tangents (e.g.\ $Q=2 k_F$ for a spherical fermi surface) because
the resulting bosonic action is singular. For further details see Refs.\
\cite{andy,ailm,hertz}. Recently, Sachdev \cite{sachdev} has questioned
whether the action used in Refs.\ \cite{andy,hertz} is adequate to describe
magnetic phase transitions in which the order parameter has O(3) or higher
symmetry because of the neglect of terms giving rise to the precession terms in
the spin-wave equation of motion, but it has been recently argued that the
action of Refs.\ \cite{andy,hertz} is in fact correct even in this case
\cite{uniform}. We do not discuss this issue here, but assume the validity of
the approach of Refs.\ \cite{andy,hertz}, which is certainly correct for low
symmetry order parameters and may apply to higher symmetries.

The partition function is
\begin{equation}
{\cal Z} = \int {\cal D}\vec\phi \, \exp{(-{\cal S}[\vec\phi])}
\end{equation}
and the action ${\cal S}[\vec\phi]$ is given by
\begin{eqnarray}\label{model}
{\cal S}[\vec\phi] &=& {\cal S}^{(0)} + {\cal S}^{(2)}[\vec\phi] +
{\cal S}^{(4)}[\vec\phi] + \dots \\ \label{gauss}
{\cal S}^{(2)}[\vec\phi] &=& V \sum_{n} \int^{\frac{\Lambda}{\xi_0}} \frac{d^d
k}{(2\pi)^d} \left[ \delta_0 + (\xi_0 k)^2 + \frac{|\Omega_n|}{\Upsilon (\xi_0
k)^{z-2}} \right] \vec\phi_n(\vec k) \vec\phi_{-n}(- \vec k) \\
{\cal S}^{(4)}[\vec\phi] &=& u_0 V^3 \sum_{n_1,\dots,n_4} \int^{\frac{\Lambda}{
\xi_0}} \frac{d^d k_1}{ (2\pi)^d} \dots \frac{d^d k_4}{(2\pi)^d}
\vec\phi_{n_1}(
\vec{k}_1) \dots \vec\phi_{n_4}(\vec{k}_4)\, \delta(\sum_{i}\vec k_i) \delta_{
\sum_{i} n_i}
\end{eqnarray}
Here ${\cal S}^{(0)}$ is the logarithm of the contribution to the partition
function from the degrees of freedom which have been integrated out. It depends
on $T$ and other external parameters. The parameter $\delta_0$ tunes the system
through the zero-temperature critical point and might be related to the
pressure
or to chemical composition. The coupling strength $u_0$ is assumed to be
constant near the critical point. The relation of $\delta_0$ to the above
mentioned control parameter $r$ will be given below. The wavevector $k$ is
measured from the
ordering wave vector $Q$. $\xi_0$ is a microscopic length, presumably of the
order of $k_{F}^{-1}$. A microscopic energy scale is given by $\Upsilon$ which
one expects to be of the order of $E_F$. The summation index $n$ labels the
Matsubara frequencies $\Omega_n = 2\pi n\,k_B\,T$. Momentum integrals are cut
off at $\frac{\Lambda}{\xi_0}$, energy integrals obtained by analytically
continuing Matsubara sums are cut off at $\Gamma\Upsilon(\xi_0 k)^{z-2}$. In a
fermi liquid one expects  the dimensionless cutoff parameters $\Lambda$ and
$\Gamma$ to be of order unity. The dots indicate terms which are irrelevant at
the critical point.

In Ref.\ \cite{andy} the critical behavior of the model was shown to be
determined by perturbative (in $u_0$) scaling about the $u_0=0$ (Gaussian)
fixed
point. The scaling procedure used in \cite{andy} maps $\cal Z$ onto a new $\cal
Z'$ identical in form to $\cal Z$ but with rescaled parameters $T(b),\delta(b)
,u(b)$. The solution of the scaling equations can be written
\begin{eqnarray}
t(b) &=& t b^z \\ \label{ureno}
u(b) &=& u b^{4-(d+z)} + {\cal O}(u^2) \\ \label{delta}
\delta(b) &=& b^2 \left\{ \delta_0 + 2 u (N+2) \int_{0}^{\log b} dx\,
e^{[2-(d+z)]x}\, f^{(2)}(t e^{z x}, \delta(e^{x})) \right\} + {\cal O}(u^2) \\
\label{freconst}
t\, {\cal S}^{(0)}(b) &=& \frac{V}{\xi_{0}^{d}} \, N \int_{0}^{\log b} dx \;
e^{-(d+z)x} \, f^{(0)}(t e^{zx}, \delta(e^x)) + S^{(0)}_0
\end{eqnarray}
Here $t=\frac{T}{T^*}$ with $T^*=\frac{\Upsilon}{k_B}$ and $u=\frac{u_0}{u^*}$
with $u^* = \frac{V^3}{\xi_{0}^{3d}}$. The analytical expressions for $f^{(0)}(
x,y)$ and $f^{(2)}(x,y)$ have been obtained in \cite{andy} and read:
\begin{eqnarray}
f^{(0)}(x, y) &=& -\frac{K_d\Lambda^{d+z-2}}{(2\pi)^d}\int_{0}^{\Gamma} \frac{
d\varepsilon}{\pi} \;\coth(\frac{\Lambda^{z-2}\varepsilon}{2 x}) \arctan(\frac{
\varepsilon}{y + \Lambda^2}) \nonumber \\ & & - \frac{2}{\pi} \Gamma \int_{0}^{
\Lambda} \frac{d^d k}{(2\pi)^d}\; k^{z-2} \coth(\Gamma\frac{k^{z-2}}{2 x})
\arctan(\frac{\Gamma}{y + k^2}) \\
f^{(2)}(x,y) &=& \frac{K_d\Lambda^{d+z-2}}{(2\pi)^d}\int_{0}^{\Gamma} \frac{
d\varepsilon}{\pi} \; \coth(\frac{\Lambda^{z-2}\varepsilon}{2 x}) \frac{
\varepsilon}{\varepsilon^2 + (y+\Lambda^2)^2} \nonumber \\ & & + \frac{2}{\pi}
\Gamma^2 \int_{0}^{\Lambda} \frac{d^d k}{(2\pi)^d}\; \coth(\Gamma \frac{k^{z-2}
}{2 x}) \frac{k^{z-2}}{\Gamma^2 + (y+ k^2)^2}
\end{eqnarray}
Note we have included the $y$-dependence of $f^{(2)}$; this was neglected in
Ref.\ \cite{andy}. Its inclusion does not affect physical results in an
important way, but makes the mathematical structure more clear.

We turn now to the solution of these equations. For this purpose, we rewrite
(\ref{delta})
\begin{equation}\label{delta-r}
\delta(b) = b^2 \big( r(t,\delta,u) - u \,\Delta(t,r(t,\delta,u),b)
\big) + {\cal O}(u^2)
\end{equation}
with
\begin{eqnarray}\label{parar}
r(t,\delta,u) &=& \delta + 2 u (N+2) \int_{0}^{\infty} dx \, e^{[2-(d+z)]x}\,
[f^{(2)}(t e^{z x}, r(t,\delta,u) \,e^{2x}) - f^{(2)}(0,0)] \\
\delta &=& \delta_0+2(N+2) u f^{(2)}(0,0)/[d+z-2]
\end{eqnarray}
and
\begin{equation}
\Delta(t,r(t,\delta,u),b) = 2(N+2)\int_{\log b}^{\infty} dx \, e^{[2-(d+z)]x}\,
f^{(2)}(t e^{zx}, r(t,\delta,u) e^{2x})
\end{equation}
We can neglect the ${\cal O}(u^2)$-terms in (\ref{delta-r}), because they do
not
affect the leading nonanalytic behavior in $u$. The expression (\ref{parar}) is
calculated in Appendix \ref{aappend}. The important result for our subsequent
considerations in this Section is that we can write $r(t,\delta,u)$ as
\begin{equation}\label{pararscale}
r(t, \delta, u) = \delta + (N+2)\,u \, t^{\frac{d+z-2}{z}}\,\phi_{d,z}\left(
\frac{r}{t^{2/z}} \right)
\end{equation}
with $\phi_{d,z}$ given by Eq.\ (\ref{phifunct}).

To calculate the free energy we use the scaling equation \cite{macri}
\begin{equation}\label{scalfree}
F(t, \delta, u) = b^{-(d+z)} F(t b^z, \delta(b), u(b)) + t\, {\cal S}^{(0)}(b)
\end{equation}
with $F(t, \delta, u)$ being the free energy of a system described by the model
(\ref{model}), calculated perturbatively in the parameter $u$ and measured in
units of $k_B T^*$. Investigation of the scale (b-) dependent terms up to
${\cal
O}(u)$ in (\ref{scalfree}) shows that we can write (\ref{scalfree}) as a
trajectory integral in phase space \cite{drnelson}:
\begin{equation}\label{traject}
F(t, \delta, u) =  \frac{V}{\xi_{0}^{d}}\, N \int_{0}^{\infty} dx \;
e^{-(d+z)x}
\, f^{(0)}(t e^{zx}, \delta(e^x)) + {\cal O}(u^2)
\end{equation}
where the ${\cal O}(u^2)$-terms do not contain the leading nonanalytic behavior
in $u$. Inserting expression (\ref{delta-r}) for $\delta(e^x)$ and performing
a variable change we get
\begin{equation}\label{unifree}
F(t, \delta, u) =  \frac{V}{\xi_{0}^{d}}\, N\, t^{\frac{d+z}{z}}\, \int_{
\frac{1}{z}\log t}^{\infty} dx \; e^{-(d+z)x} \, f^{(0)}\left( e^{zx},
\frac{r}{
t^{\frac{2}{z}}} \, \left[ 1 - \frac{u t^{\frac{d+z-2}{z}}}{r}\, \Delta\left(
1, \frac{r}{t^{\frac{2}{z}}}, e^x \right)\right]\, e^{2 x}\right)
\end{equation}
It is therefore apparent that we can write the free energy (apart from
non-singular contributions) as
\begin{equation}
F(t,\delta,u) = t^{\frac{d+z}{z}}\, f(\frac{r}{t^{\frac{2}{z}}}, \frac{u t^{
\frac{d+z-2}{z}}}{r})
\end{equation}
Due to the relation (\ref{pararscale}) and the definition of $\Delta(t,r(t,
\delta,u),b)$, this scaling behavior is preserved when taking derivatives
with respect to temperature, and we find the specific heat coefficient to obey
(\ref{singpart}). It is shown in Appendix \ref{cappend} that the dependence on
the variable $\frac{u t^{\frac{d+z-2}{z}}}{r}$ leads to negligible corrections
to the leading scaling behavior. The latter is given by
\begin{equation}\label{leading}
\gamma_{leading} = \frac{V}{\xi_{0}^{d}}\, N \int_{0}^{\infty} dx \; e^{(z-d)x}
\,\partial_{\zeta}^2 f^{(0)}(\zeta, r e^{2x})\Big|_{\zeta = t e^{z x}}
\end{equation}
and is evaluated in detail in Appendix \ref{bappend}.

In order to characterize the specific heat near the transition line
$t_c(\delta)$ we calculate the discontinuity (jump) of the specific heat as
it is predicted by mean field theory, using the scaling equation
(\ref{scalfree}) for the free energy. Very close to the true transition, the
model (\ref{model}) can be scaled to a classical Gaussian model given by the
Landau-Ginzburg functional \cite{andy}
\begin{eqnarray}
\exp{(-\frac{F(t(b), \delta(b), u(b))}{t(b)})} &=& \int \mbox{$\cal D$}\phi \,
\exp{(-\mbox{$\cal S$}[\vec\phi])} \\ \mbox{$\cal S$}[\vec\phi] &=&
\int d^d x\, \left\{\frac{1}{2} |\vec\nabla \vec\phi|^2 + \delta(b)
|\vec\phi|^2 + v(b) (|\vec\phi|^2)^2\right\} \label{classmodel}
\end{eqnarray}
with $v(b)=u(b)t(b)$. Applying the usual mean field approximation yields the
discontinuity of the specific heat. Analytical details of this calculation are
given in Appendix \ref{dappend}. The result is Eq. \ref{3Djump}. The correction
to the specific heat jump due to
Gaussian fluctuations is also evaluated in this Appendix.  The result
is Eq. (\ref{3Ddiverge}). Non-gaussian critical fluctuations become important
when the Ginzburg criterion is violated \cite{macri}.   The Ginzburg criterion
may be obtained by demanding that the Gaussian fluctuation contribution to the
specific heat is small compared to the mean field specific heat jump.
The result is $u t_c r^{\frac{d-4}{2}}\ll 1$.  This is violated for $t-t_c <
t_{
c}^{2-\frac{1}{z}}$ in 3 dimensions.  The expressions for the Ginzburg
criterion and the width about the transition line where it is violated correct
the results obtained in \cite{andy}, which were in error by factors of
$t_c^{1/z
}$. As the critical temperature is small, mean field behavior characterizes the
transition correctly except from a negligibly small region about $t_c$.

\section{Results for 3D}\label{results}
We give the results for the specific heat coefficient (normalized per mol) and
for the transition temperature in the physically relevant case of 3 dimensions
in this section.

The formula describing the specific heat in the whole Gaussian region of the
phase diagram is Eq.\ (\ref{gaussheat2}) [which can be simplified to
(\ref{gaussd3z2}) for $z=2$]. However, in order to compare experimental data
to the prediction of this formula it is necessary to extract the
system-dependent scale $T^*$ and the scale factor which relates $r(T=0)$ to
experimentally accessible control parameters like pressure or concentration
first. For this purpose it is useful to have expressions for $\gamma$ in the
quantum ($T\ll T^* r$) and classical ($T\gg T^* r$) limits in the phase
diagram.
Fits of the data to the appropriate functional forms provide the scales  which
can then be inserted into (\ref{gaussheat2}), which must in general be
evaluated
numerically.

The scale factors $r^*$ (which we have set equal to 1) and $T^*$ are of course
arbitrary.  Only dimensionless combinations of the coefficients we present are
universal. However, there is a natural definition of the scale factor which
gives an estimate of the important microscopic scales of the system.  We have
defined  $T^*$  so that for $d=3$ and $z=2$ the leading low-$T$ non-singular
contribution to the Gaussian model specific heat is
$\frac{k_B}{T^*}\frac{\Lambda}{
\xi_{0}^{d}}\frac{N}{3\pi}$. For a density of order 1,
$\xi_{0}^{-1}\sim k_F \sim (3\pi^2)^\frac{1}{3}$ so the specific heat
coefficient would be of the order of the leading fermi-liquid term in
(\ref{norfersh}).

Furthermore, it is important to note that the specific heat calculated in this
communication is only the contribution arising from the existance of a quantum
critical point. A metallic system has an additional background specific heat
given by (\ref{norfersh}). Depending on the dimensionality and the dynamical
exponent, the leading behavior or corrections to the leading behavior in
(\ref{norfersh}) are changed due to critical fluctuations.

{\bf Antiferromagnetism ($z=2$):}
The leading term of $\gamma$ is still the normal fermi liquid contribution, a
nonuniversal constant $\gamma_0$. Only corrections to this leading behavior are
affected by the critical point. In the quantum regime ($T\ll T^* r$) we find:
\begin{equation}\label{start}
\gamma^{quantum, z=2} = \gamma_0 - \frac{k_B N_A}{T^*}\, \frac{N}{6}\,
r^{\frac{
1}{2}} - \frac{k_B N_A}{(T^*)^3 }\, \frac{N \pi^2}{60}\,
\frac{T^2}{r^{\frac{3}{
2}}}\, +\,\dots
\end{equation}
Thus the coefficient of the first (${\cal O}(T^2)$) correction to the fermi
liquid behavior diverges as $r\rightarrow 0$ in the quantum regime. The
classical limit ($T\gg T^* r$) yields
\begin{equation}\label{classgauss}
\gamma^{classical, z=2} = \gamma_0 - \frac{k_B
N_A}{(T^*)^\frac{3}{2}}\,\frac{15
N}{64}\left(\frac{2}{\pi}\right)^{\frac{3}{2}}\,\zeta(\frac{5}{2})\, T^{\frac{1
}{2}} - \frac{k_B N_A}{(T^*)^\frac{1}{2}}\, \frac{3
N}{124}\,\left(\frac{2}{\pi}
\right)^{\frac{3}{2}}\, \zeta(\frac{3}{2}) \,\frac{r}{T^{\frac{1}{2}}} + \dots
\end{equation}
with $\zeta(\frac{5}{2})=1.34149$ and $\zeta(\frac{3}{2})=2.61238$. The term
linear in $r$ dominates corresponding terms in the non-critical part of the
specific heat because it is enhanced by a factor $T^{-\frac{1}{2}}$ at low
temperatures.

{\bf Clean Ferromagnet ($z=3$):}
In the quantum regime both the leading (constant) term and the
$T^2\log T$-correction of the normal fermi liquid expression (\ref{norfersh})
are enhanced by critical fluctuations:
\begin{equation}
\gamma^{quantum,z=3} = \gamma_0 - \frac{k_B N_A}{T^*}\, \frac{N}{6\pi}\,\log r
- \frac{k_B N_A}{(T^*)^3}\, \frac{2\pi N}{15}\, \,\frac{T^2}{r^3}\log \left(
\frac{(T/T^* )^2}{r^3}\right) +\dots
\end{equation}
The classical regime shows non-fermi-liquid behavior:
\begin{equation}
\gamma^{classical,z=3} = - \frac{k_B N_A}{T^*}\, \frac{N}{9\pi}\, \log \left(
\frac{T}{\tilde T}\right) - \frac{k_B N_A}{(T^*)^\frac{1}{3}}\, \frac{N \Gamma(
\frac{7}{3}) \zeta(\frac{4}{3})}{12\pi^2\sqrt{3}}\, \frac{r}{T^{\frac{2}{3}}} +
\dots
\end{equation}
with $\Gamma(\frac{7}{3})\zeta(\frac{4}{3})=4.28742$.

{\bf Dirty Ferromagnet ($z=4$):}
In this case, the leading behavior in the quantum regime is of fermi liquid
type
and enhanced by critical fluctuations, but the first correction is not fermi
liquid like, and the coefficient of the latter diverges when approaching the
critical point.
\begin{equation}
\gamma^{quantum,z=4} = \frac{k_B N_A}{T^*}\,\frac{N}{6}\,\,r^{-\frac{1}{2}} -
\frac{k_B N_A}{(T^*)^\frac{3}{2}}\, \frac{15 N}{64}\,\left(\frac{2}{\pi}
\right)^{\frac{3}{2}}\,\zeta(\frac{5}{2})\,\frac{T^{\frac{1}{2}}}{r^{\frac{3}{2}
}} + \dots
\end{equation}
In the classical regime, the leading behavior is determined by critical
fluctuations and is not of fermi liquid type:
\begin{equation}\label{finish}
\gamma^{classical,z=4} = - \frac{k_B N_A}{(T^*)^\frac{3}{4}}\, \frac{N}{8\pi^2
}\,\frac{\Gamma(\frac{11}{4}) \zeta(\frac{7}{4})}{\sin(\frac{7}{8}\pi)}\,T^{-
\frac{1}{4}} - \frac{k_B N_A}{(T^*)^\frac{1}{4}}\, \frac{N}{32\pi^2}\, \frac{
\Gamma(\frac{9}{4})\zeta(\frac{5}{4})}{\sin(\frac{3}{8}\pi)}\,\frac{r}{T^{
\frac{3}{4}}} + \dots
\end{equation}
with $\Gamma(\frac{11}{4})\zeta(\frac{7}{4})=3.15612$ and $\Gamma(\frac{9}{4})
\zeta(\frac{5}{4})=5.20628$.

{\bf Transition region:} In order to characterize the region near the
transition
line $T_c(\delta)$ we give the result for this curve in 3D
\begin{equation}\label{3Dtransi}
T_c = T^* \,\left[ \frac{-\delta}{{\cal D}_{3,z}\, (N+2)\, u}
\right]^\frac{z}{1+z}
\end{equation}
the expression for the specific heat jump predicted by mean field theory
\begin{equation}\label{3Djump}
\Delta C = \frac{k_B N_A}{(T^*)^{\frac{2}{z}(z+1)}} N(N+2)^2 \,
{\cal D}_{3,z}^{2} \, u T_c^{\frac{2+z}{z}}
\end{equation}
and the correction to the mean field result due to Gaussian fluctuations
\begin{equation}\label{3Ddiverge}
C^{transition} = k_B N_A \,\frac{N}{16\pi}\,\left(\frac{z}{z+1}\right)^{\frac{1
}{2}}\,\left[ {\cal D}_{3,z} (N+2)
u\left(\frac{T_c}{T^*}\right)^{\frac{z+1}{z}}
\right]^{\frac{3}{2}}\,\left[\frac{T - T_c}{T_c}\right]^{-\frac{1}{2}}
\end{equation}
The numerical values for ${\cal D}_{3,z}$ are ${\cal D}_{3,2}=0.52109$, ${\cal
D}_{3,3} =0.39396$, and ${\cal D}_{3,4}=0.35875$.

\section{Comparison to experimental data}\label{experi}
The discovery of superconductivity in so-called heavy-fermion systems
\cite{review} resulted in extensive studies of these materials. Some of them
undergo magnetic phase transitions at low temperatures, others display no
long-range order. It has become evident, however, that several of the latter
are
near magnetic critical points as well. As an example, modest doping of CeCu$_6$
\cite{schlag,germ} and UPt$_3$ \cite{visser1,visser2} leads to
antiferromagnetism. The influence of pressure variation on the critical
behavior
has been studied as well \cite{germ,brodale}. The reason for the proximity of
the heavy-fermion metals to a quantum critical point is the extreme sensitivity
of the two competing energy scales (fermi [Kondo] temperature for the fermi
liquid state and RKKY exchange energy for the magnetically ordered state) in
these materials to changes in parameters such as the lattice constant.
Therefore
the heavy fermion compounds may be an appropriate experimental system to study
the implications of the theory presented in this communication.

{\bf Application to CeCu$_2$Si$_2$}

Measurements of the temperature-dependence of the specific heat have been
carried out at compounds Ce$_{1-x}$M$_x$Cu$_{2}$Si$_2$ with M=La,Y,Lu and
$x=0.0,0.05,0.1,0.2,0.5\mbox{ and }0.9$ \cite{csjee}. Doping with La expands
the
lattice, whereas doping with either Y or Lu contracts it. As the data for
La-doped samples show a rather complicated behavior, we will not comment on
them
in this work. Diluting CeCu$_{2}$Si$_2$ with Y,Lu indicates, however, that this
material is near a quantum-critical point. Based on the assumption that this is
an antiferromagnetic instability, we present an interpretation of the data for
Ce$_{1-x}$Lu$_x$Cu$_{2}$Si$_2$. The numbers and conclusions which can be drawn
from the experiments carried out with Y-doped samples do not differ
significantly from those given below.

In Fig.\ref{comp} the data for various concentrations $x$ are shown. The
parameter tuning the system through the quantum critical point $\delta=r(T=0)$
is defined as $\delta=\frac{x-x_c}{x^*}$. If the pure compound is right at the
critical point ($x_c=0$), this system would be in the classical regime and we
would expect to find the low-temperature specific heat behaving according to
(\ref{classgauss}). Indeed the low-temperature ($T\le 5$K) data for $x=0$ fit
nicely to the square root temperature-dependence. This fit yields the
temperature scale $T^*=5K$ which is of the order of the Kondo temperature in
this material: $T_K=10K$ \cite{steglich}. Furthermore, the first correction to
the leading behavior in the classical regime is linear in $r$ and therefore
linear in $\delta$ as well. As we will check below, the compounds for $x=0.05,
0.1$ are still in the classical regime. We take the measured values for
$\frac{C
}{T}(T=1.1$K) given in \cite{csjee} at concentrations $x=0.0$ and $x=0.05$ in
order to determine the scale $x^*$. We find $x^*=0.3$, hence for the
quantum-classical crossover temperature $T_{cross}=\frac{T^*}{x^*}\, x = 17 x$
which is reached right at the beginning of the measurements for $x=0.05,0.1$,
whereas for $x=0.2$ the quantum-classical crossover is accessed experimentally.
The data for $x=0.5,0.9$ and $T>5K$ cannot be explained within the framework of
the theory presented here because the systems are already too far away from the
critical point. Taking the values for the nonuniversal scale factors $T^*=5$K,
$x^*=0.3$, we can apply (\ref{gaussd3z2}) to predict the temperature dependence
of the specific heat for $x=0.05,0.1,0.2$. As Fig.\ref{comp} shows, theory and
data agree quite well for low temperatures. Please note that we neglected any
effect of the dangerously irrelevant variable $u$. The coincidence for $x=0.2$
at low temperatures is particularly nice because there the system is in the
quantum regime, whereas we extracted all information for the theory from the
classical regime. There is apparently a systematic deviation from the theory
occurring at lower temperatures for larger $x$. We do not understand this
deviation at present. We suspect that the deviations (starting at lower
temperatures for increasing concentration) are due to noncritical fluctuations
which become important farther away from the critical point, but the observed
interplay between $T$ and $x$ is not expected in our model.

\section{Conclusion}\label{conclusion}
We have determined the universal functions needed to quantitatively analyze
the behavior of the specific heat near a quantum critical point with dynamical
exponent $z=2,3,4$ in $d=3$ spatial dimensions and illustrated, by an example,
the use of the equations. The results are summarized in Eqs.\
(\ref{start})--(\ref{3Ddiverge}), (\ref{gaussheat2}) and (\ref{gaussd3z2}),
respectively. Depending on the dynamical exponent $z$, the leading or
subleading
behavior of the normal fermi liquid specific heat coefficient is affected due
to
the proximity to the quantum critical point.

The physical properties of these transitions are governed by four non-universal
scales: a temperature scale $T^*$, a scale $r^*$ for the Hamiltonian parameter
tuning the $T=0$ transition, a dangerously irrelevant variable $u$, and a
length
scale $\xi_0$. Once these are known, all the behavior is fixed. The specific
heat does not involve the length scale $\xi_0$ but provides sufficient
information to overdetermine the remaining parameters.

An important open theoretical problem is to understand the nature and magnitude
of the corrections to the universal behavior. On the experimental side, it
would be useful to study in detail the specific heat of a system to which the
theory applies.

\acknowledgments

We thank P.\ C.\ Hohenberg for helpful discussions.
U.\ Z.\ was supported in part by the Studienstiftung des deutschen Volkes
(German National Scholarship Foundation).

\newpage

\appendix
\section{Quantum parameter and critical temperature}\label{aappend}
In this appendix we will calculate (\ref{parar}) and obtain an expression for
the critical temperature.

We first extract the explicit temperature dependence of $r$ when writing
\begin{eqnarray}
r(t, \delta, u) &=& \tilde r(r(t,\delta,u), u) + \tilde{\tilde r}(t,
r(t,\delta,u), u) \\
\tilde r(\delta, u) &=& \delta + 2 u (N+2)\int_{0}^{\infty} dx \,
e^{[2-(d+z)]x}
\, \left[f^{(2)}(0, r(t,\delta,u) \,e^{2x}) - f^{(2)}(0,0)\right]
\label{parar1}
\\
\tilde{\tilde r}(t, \delta, u) &=& 2u(N+2)\int_{0}^{\infty} dx\, e^{[2-(d+z)]x}
\, \left[ f^{(2)}(t e^{z x}, r(t,\delta,u) \,e^{2x}) - f^{(2)}(0, r(t,\delta,u)
\, e^{2x}) \right] \label{parar2}
\end{eqnarray}
Further inspection of Eq.\ (\ref{parar1}) reveals that
\begin{equation}
\tilde r(\delta, u) = \delta\, \left( 1 + 2 u (N+2) \frac{\partial_{\eta}
f^{(2)}(0, \eta)\big|_{\eta=0}}{d+z-4} \right) +  2 u (N+2) \left\{
\begin{array}{ll}
{\cal O}(r^{\frac{d+z-2}{2}}) &\mbox{ if } d+z-6\ne 0 \\ {\cal O}(r^2 \log r)
& \mbox{ if } d+z-6=0
\end{array}
\right.
\end{equation}
where we omitted ${\cal O}(u^2)$-terms like those we have already neglected
when
writing (\ref{delta-r}). The dangerously irrelevant variable $u$ leads to a
rescaling of $\delta$ by a $u$-dependent factor, which can be absorbed into the
scale (unit of measurement) of $\delta$. Note that the nonanalytic
contributions
in $r$ due to the variable $u$ are nonleading and therefore negligible near
the critical point.

Neglecting nonleading (analytic) contributions we can write for (\ref{parar2})
\begin{equation}
\tilde{\tilde r}(t, \delta, u) = u (N+2) \, t^{\frac{d+z-2}{z}}\, \phi_{d,z}
\left( \frac{r}{t^{2/z}}\right) + {\cal O}(u t^2)
\end{equation}
with
\begin{equation}\label{phifunct}
\phi_{d,z}(x) = \frac{K_d}{(2\pi)^{d+1}} 8 \int_{0}^{\infty} dy\,
y^{\frac{z-d}{
2}} \, \int_{0}^{\infty} d\varepsilon\, \frac{\varepsilon (\coth\varepsilon -
1)
}{4 \varepsilon^2 y^z + (x\, y + 1)^2}
\end{equation}
The two limits $\frac{r}{t^{2/z}}\ll 1$ and $\frac{r}{t^{2/z}}\gg 1$ correspond
to the classical and quantum regimes, respectively \cite{andy}. We find
$$ \begin{array}{ll}
\phi_{d,z}(x) \Rightarrow {\cal D}_{d,z} & \mbox{ for } x\rightarrow 0 \mbox{
(classical) } \\
\phi_{d,z}(x) \Rightarrow x^{\frac{d-z-2}{2}} & \mbox{ for } x\rightarrow\infty
\mbox{ (quantum) }
\end{array}$$
with
\begin{equation}\label{dconstant}
{\cal D}_{d,z} = \frac{K_d}{(2\pi)^d}\, \frac{2}{z} \frac{\Gamma(\frac{d-2}{z}
+
1) \zeta(\frac{d-2}{z}+1)}{\sin \frac{z+2-d}{2z}\pi}
\end{equation}
Numerical values for ${\cal D}_{d,z}$ in three dimensions and for the dynamical
exponents considered in this communication are ${\cal D}_{3,2}=0.52109$, ${\cal
D}_{3,3} =0.39396$, and ${\cal D}_{3,4}=0.35875$. The term with explicit
temperature dependence in the quantum regime is $t^2 r^{\frac{d-z-2}{2}}$ which
is small compared to the contribution $r^\frac{d+z-2}{2}$ to $\tilde r$ which
we
have already neglected. The general result is therefore
\begin{eqnarray}
r(t, \delta, u) &=& \delta + (N+2)\,u \, t^{\frac{d+z-2}{z}}\,\phi_{d,z}\left(
\frac{r}{t^{2/z}} \right) \\
&=& \delta \mbox{ in the quantum regime} \\ \label{transi}
&=& \delta + {\cal D}_{d,z}\, (N+2)\, u\, t^{\frac{d+z-2}{z}} \mbox{ in the
classical regime}
\end{eqnarray}

The true transition occures when $r$ vanishes. Therefore we find an expression
for the transition temperature $t_c$ by setting (\ref{transi}) equal to zero
and solving for $t$:
\begin{equation}
t_c = \left[ \frac{- \delta}{ {\cal D}_{d,z}\, (N+2)\, u}
\right]^{\frac{z}{d+z-2}}
\end{equation}

\section{Quantum-classical crossover}\label{bappend}
In this section we investigate the expression (\ref{leading}) for the leading
contributions to the specific heat coefficient, which arise from the
quantum-classical crossover.

The expression (\ref{leading}) can be written
\begin{eqnarray}
\gamma_{leading} &=& \frac{V}{\xi_{0}^{d}}\,\frac{4 K_d N}{(2\pi)^{d+1}} \int_{
\frac{1}{\Lambda^2}}^{\infty} dx \; x^{\frac{z-d}{2} - 1}\int_{0}^{\infty}
d\varepsilon\, \frac{\varepsilon^2}{\sinh^2\varepsilon}\frac{r x + 1}{(2 t
\varepsilon)^2\,x^z + (r x + 1)^2} \\ \label{gaussheat1}
&=&  r^{\frac{d-z}{2}}\, \frac{V}{\xi_{0}^{d}}\,\frac{4 K_d N}{(2\pi)^{d+1}}
\int_{0}^{\infty} d\varepsilon\, \frac{\varepsilon^2}{\sinh^2\varepsilon}\int_{
\frac{r}{\Lambda^2}}^{\infty} dx\, x^{\frac{z-d}{2} - 1} \frac{x+1}{(x+1)^2 +
\frac{(2 t \varepsilon)^2}{r^z} \, x^z} \\ \label{gaussheat2}
&=& t^{\frac{d-z}{z}}\, \frac{V}{\xi_{0}^{d}}\,\frac{2^{\frac{d+z}{z}} K_d N}{
(2 \pi)^{d+1}} \, \int_{0}^{\infty} d\varepsilon\, \frac{\varepsilon^{\frac{d+z
}{z}}}{\sinh^2\varepsilon}\, \int_{\frac{(2 t \varepsilon)^{
2/z}}{\Lambda^2}}^{
\infty} dx\, x^{\frac{z-d}{2} - 1} \frac{\frac{r}{(2 t \varepsilon)^{2/z} }\, x
+ 1}{\left( \frac{r}{(2 t \varepsilon)^{2/z}}\, x + 1\right)^2 + x^z}
\end{eqnarray}
In the case of $d\ne z$, the dependence on the cut-off $\Lambda$ gives rise to
analytic terms ${\cal O}(r, t^2)$ only, which we will neglect in the following.
To discuss the case $d=z$, however, the existance of a lower boundary of the
$x$-integral is essential for technical reasons because logarithms occure. But
even in this case, the nonanalytic behavior of the heat capacity does not
depend
on the cut-off.

For $d=3,z=2$ it is possible to perform the $x$-integrals and we find for the
singular part of $\gamma$ which arises from the quantum-classical crossover:
\begin{equation}\label{gaussd3z2}
\gamma_{leading}^{d=3,z=2} = - t^{\frac{1}{2}}\, \frac{V}{\xi_{0}^{3}}\,
\frac{N}{\sqrt{2}\pi^2} \, \int_{0}^{\infty}d\varepsilon\,
\frac{\varepsilon^2}{
\sinh^2 \varepsilon} \sqrt{\frac{r}{t} + \sqrt{\left(\frac{r}{t}\right)^2 + (2
\varepsilon)^2}}
\end{equation}

In the next paragraph we study the asymptotic behavior of $\gamma_{leading}$ in
the classical and quantum regimes corresponding to the limits $\frac{r}{t^{2/z}
}\ll 1$ and $\frac{r}{t^{2/z}}\gg 1$, respectively. For sake of simplicity,
we do not write the factor $\frac{V}{\xi_{0}^{d}}\frac{K_d N}{(2\pi)^{d+1}}$
explicitly. Results of the following paragraphs have to be multiplied by this
term to obtain the correct results.

{\bf Classical regime:}
To investigate the classical limit it is most convenient to use Eq.\
(\ref{gaussheat2}) and expand in the parameter $\frac{r}{t^{2/z}}$ (which is
small in the classical regime). The leading behavior is
\begin{equation}
\gamma_{leading}^{classical, d\ne z} = \frac{2\pi}{z}\frac{\Gamma(\frac{d}{z} +
2) \zeta(\frac{d}{z} + 1)}{\sin\left(\frac{d+z}{2z}\pi\right)} \,\, t^{\frac{
d-z}{z}} + \dots
\end{equation}
in the case $d\ne z$, whereas for $d=z$ we have
\begin{equation}
\gamma_{leading}^{classical, d=z} = - \frac{4 \pi^2}{3 d}\,\log t + \dots
\end{equation}
The first correction to the leading term is obtained by an expansion in the
small parameter and reads
\begin{equation}
-\frac{\pi}{2 z} \frac{\Gamma(\frac{d-2}{z} + 2)\zeta(\frac{d-2}{z} + 1)}{
\sin\left(\frac{2+z-d}{2z}\pi\right)} \, \frac{r}{t^{\frac{2+z-d}{z}}}
\end{equation}

{\bf Quantum regime:}
In this limit we use an expansion of (\ref{gaussheat1}) in the small parameter
$\frac{t^2}{r^z}$. The leading term for $d\ne z$ is
\begin{equation}
\gamma_{leading}^{quantum, d\ne z} = \frac{2}{3} \pi^3 \mbox{ sgn}(z-d) \, r^{
\frac{d-z}{2}} + \dots
\end{equation}
In the case $d=z$ we find
\begin{equation}
\gamma_{leading}^{quantum, d=z} = -\frac{2}{3} \pi^2 \, \log
\frac{r}{\Lambda^2}
+ \dots
\end{equation}
For $z=2$ the corrections to the leading behavior are obtained by expanding
(\ref{gaussheat1}) in $\frac{t^2}{r^z}$. This yields the asymptotic series
\begin{eqnarray}
\gamma_{leading}^{quantum, d\ne z=2} &=& - 4\pi^3 \, r^{\frac{1}{2}}\,\sum_{n=0
}^{\infty} (-1)^n (2\pi)^n |B_{2n+2}|\frac{\Gamma(2n - \frac{1}{2})}{\Gamma(-
\frac{1}{2})\Gamma(2n+1)}\, \left(\frac{t}{r}\right)^{2n} \\ \label{quant}
&=& -\frac{2\pi^3}{3}\, r^{\frac{1}{2}} - \frac{\pi^5}{15}\,\frac{t^2}{r^{
\frac{3}{2}}}\,+\,\dots
\end{eqnarray}
where $B_m$ denote Bernoulli numbers.
Such an expansion is inappropriate to find corrections in the cases $z=3,4$
because these are nonanalytic in the quantity $\frac{t^2}{r^z}$. It is more
convenient to rewrite (\ref{gaussheat2}) for large $\frac{r}{t^{2/z}}$ and
subtract the already calculated leading term. The corrections are then given by
\begin{equation}
-\frac{2^{\frac{d}{z-2} + 1}}{z-2}\,\frac{t^{\frac{d}{z-2} -1}}{r^{\frac{d}{z
-2}}} \int_{0}^{\infty}d\varepsilon\, \frac{\varepsilon^{\frac{d}{z-2} + 1}}{
\sinh^2 \varepsilon} \,\int_{\frac{(2 t \varepsilon)^{2/z}}{r^z}}^{\infty} dx\,
\frac{x^{\frac{z-d-2}{2(z-2)}}}{1+x}
\end{equation}
which gives for $d=z=3$
\begin{equation}
\frac{8}{15}\pi^4\,\frac{t^2}{r^3}\log \frac{t^2}{r^3} + \dots
\end{equation}
and $d=3, z=4$
\begin{equation}
-\frac{15}{2}\left(\frac{\pi}{2}\right)^{\frac{3}{2}}\zeta(\frac{5}{2})\,
\frac{t^\frac{1}{2}}{r^{\frac{3}{2}}} +\dots
\end{equation}

\section{Dangerously irrelevant variable and the quantum-classical crossover}
\label{cappend}
In this appendix we investigate the dependence of (\ref{unifree}) on the
argument $\frac{u t^{\frac{d+z-2}{z}}}{r}$ in the quantum and classical
regimes.
We show that it leads to corrections to the leading behavior given by
(\ref{singpart}), which is calculated in Appendix \ref{bappend}.

In the quantum regime ($t\ll r^{z/2}$) we can write (\ref{unifree}) as
\begin{equation}
F(t, \delta, u) =  \frac{V}{\xi_{0}^{d}}\, N\, r^{\frac{d+z}{2}}\, \int_{
\frac{1}{2}\log r}^{\infty} dx \; e^{-(d+z)x} \, f^{(0)}\left(
\frac{t}{r^{z/2}}
e^{zx}, \left[ 1 - u r^{\frac{d+z-4}{2}}\, \Delta\left(\frac{t}{r^{z/2}}, 1,
e^x
\right)\right]\, e^{2 x}\right)
\end{equation}
As the parameter $\frac{t}{r^{z/2}}$ is small, an expansion in this quantity is
reasonable. The term $u r^{\frac{d+z-4}{2}}$ is small compared to unity near
the
critical point ($u$ is assumed to be of order unity, and $r$ is small), it
leads
to corrections to coefficients in the leading behavior in the quantum regime,
the latter is given by an expansion of (\ref{gaussheat1}) in the small
parameter
$\frac{t}{r^{z/2}}$.

Similarly, in the classical regime we can expand (\ref{unifree}) in the small
quantity $\frac{r}{t^{2/z}}$. The correction due to the argument $\frac{u t^{
\frac{d+z-2}{z}}}{r}$ is here $u t^{\frac{d+z-4}{z}}$, which again has to be
compared to unity. Near the critical point, this correction is therefore
negligible. However, a careful calculation of the terms arising when taking
derivatives w.r.t.\ temperature shows that a nonanalyticity arises due to the
temperature dependence of $r(t,\delta,u)$ in the classical regime. This leads
to a divergence of the specific heat which is calculated in Appendix
\ref{dappend}.

In conclusion we find the variable $u$ to give rise to corrections to
coefficents in the limiting expansions of $\gamma_{leading}$ (cf.
(\ref{gaussheat1}) and (\ref{gaussheat2}), respectively) only -- apart from a
narrow region about the transition line $t_c(\delta)$, where the Gaussian
correction to the discontinuity of the specific heat becomes important (see
discussion in the last paragraph of Section \ref{formal}). The coefficients in
the expansion of $\gamma_{leading}$ are of order unity, whereas the corrections
in $u r^{\frac{d+z-4}{2}}$ and $u t^{\frac{d+z-4}{z}}$ are small near the
critical point.

\section{Specific heat jump and divergence near the true transition}
\label{dappend}
The characteristic behavior of the model (\ref{model}) near the transition is
calculated in this section. As it has been shown in \cite{andy}, (\ref{model})
scales to a classical theory given by (\ref{classmodel}). For this theory,
the discontinuity of the free energy reads \cite{macri}
\begin{equation}
\Delta F(b) = t(b)\frac{\delta(b)^2}{4 v(b)} = \frac{\delta(b)^2}{4 u(b)}
\end{equation}
Using the relation $F(1) = b^{-(d+z)}\, F(b)$ in order to match the scaled
model
to the original one, we find the expression for the discontinuity of the free
energy of model (\ref{model}). We insert (\ref{ureno}) for $u(b)$ and
substitute
$\delta(b)= b^2\, r(t,\delta,u)$ with $r(t,\delta,u)$ given in Eq.\
(\ref{transi}). The latter is justified according to (\ref{delta-r}), noting
the
fact that the quantity $\Delta(t,\delta,u)$ becomes negligible in the classical
regime. Differentiating twice w.r.t.\ temperature and taking into account that
$r(t,\delta,u)$ vanishes at the transition, one ends up with the quantitative
expression for the specific heat jump:
\begin{equation}\label{jump}
\Delta C = {\cal D}^{2}_{d,z}\, N(N+2)^2\, u\, t_{c}^{1+\frac{2}{z} (d-2)}
\end{equation}
with the numerical constant ${\cal D}_{d,z}$ given by Eq.\ (\ref{dconstant}).
Please note the anomalous exponent of $t_c$ in this expression. In theories
where the distance to the critical point is measured by $(t-t_c)$ -- classical
Landau theories -- we find that $\Delta C$ is proportional to $t_c$. For our
case of a quantum critical point, the specific heat jump turns out to be
smaller
because additional (in general noninteger) powers of $t_c$ occure.

Although the effective dimensionality of the theory (\ref{model}) is $d+z$
(which is greater than 4 in the cases discussed in this communication), a
divergence of the specific heat occures very near the true transition (for
vanishing $r(t,\delta,u)$). This is due to the fact that the quantum
fluctuations become irrelevant and a classical description of the transition
becomes more and more appropriate when the critical temperature is approached.
To calculate the divergence, which has its origin in the temperature dependence
of $r(t,\delta,u)$, we start from (\ref{unifree}) and express the temperature
derivatives (which we have to take in order to obtain $\gamma$) in terms of
derivatives w.r.t. $r(t,\delta,u)$. Apart from contributions which are
negligible corrections like those calculated in Appendix \ref{cappend}, and
analytic terms, we get
\begin{equation}
\gamma^{transition} = (\partial_{t} r)^2\, \frac{2 K_d N}{(2\pi)^{d+1}}\, r^{
\frac{d+z-4}{2}}\,\int_{0}^{\infty} dy\, y^{\frac{4-d-z}{2} - 1}\, \int_{0}^{
\infty} d\varepsilon\, \coth(\frac{\varepsilon}{2 \frac{t}{r^{z/2}}
y^{\frac{z}{
2}}}) \frac{\varepsilon\, (y+1)}{[\varepsilon^2 + (y + 1)^2]^2}
\end{equation}
The divergence occures for $r(t,\delta,u)\rightarrow 0$, we can therefore
expand
the hyperbolic cotangent in its argument. Only the leading term gives rise to a
divergence in $\gamma^{transition}$, which is
\begin{equation}
\gamma^{transition} = (\partial_t r)^2\, \frac{4 K_d N}{(2\pi)^{d+1}}\, t\, r^{
\frac{d-4}{2}}\, \int_{0}^{\infty} dy\, y^{\frac{4-d}{2}-1}\, (y+1)\,\int_{0}^{
\infty} d\varepsilon \,\frac{1}{[\varepsilon^2 + (y+1)^2]^2}
\end{equation}
Performing the (elementary) integrals and using (\ref{transi}) to evaluate
$\partial_t r$ yields finally:
\begin{equation}\label{divergent}
C^{transition} = \frac{N K_d (d-2)}{(2\pi)^{d-1} 16 \sin\frac{d-2}{2}\pi}\left(
\frac{d+z-2}{z}\right)^{\frac{d-4}{2}}\, \left( {\cal D}_{d,z}\,(N+2)\, u\,
t_{c}^{\frac{d+z-2}{z}} \right)^\frac{d}{2} \left[ \frac{t-t_c}{t_c}
\right]^{\frac{d-4}{2}}
\end{equation}
The numerical constant ${\cal D}_{d,z}$ is given by Eq.\ (\ref{dconstant}).

\begin{figure}
\caption{Phase diagram for a quantum-critical point which is controled by a
zero temperature
parameter $\delta$. The shaded region is the (magnetically) ordered phase.
The solid line denotes the transition curve $T_c(\delta)$, the dotted line
corresponds to the quantum-classical crossover which occures when $T$ becomes
of the order of $r^{z/2}$.}
\label{diagram}
\end{figure}

\begin{figure}
\caption{Comparison of experimental data measured from CeCu$_2$Si$_2$ to the
theoretical prediction for $d=3,z=2$. The temperature scale $T^*=5$K has been
extracted from the data for $x=0$, assuming the pure compound being at the
critical point. The scale $x^*=0.3$ is determined by the knowledge of $T^*$ and
the datapoint at $x=0.05,T=1.1$K. Solid curves represent the theoretical
prediction, symbols are data points.}
\label{comp}
\end{figure}

\end{document}